\documentclass[twocolumn]{revtex4-2}
\usepackage{graphicx}
\usepackage{amsmath}
\usepackage{verbatim}
\usepackage{hyperref}
\usepackage{amsfonts}
\usepackage{adjustbox}

\usepackage[normalem]{ulem}
\usepackage{color,soul}

\begin{document}

\title{Magneto-Crystalline Composite Topological Defects and Half-Hopfions}

\author{Sahal Kaushik}
\affiliation{Nordita, Stockholm University and KTH Royal Institute of Technology, Hannes Alfv\'{e}ns v\"{a}g 12, Stockholm, SE-106 91,  Sweden}
\email{sahal.kaushik@su.se}
\author{Filipp N. Rybakov}
\affiliation{Department of Physics and Astronomy, Uppsala University, Box-516, Uppsala SE-751 20, Sweden}
\author{Egor Babaev}
\affiliation{Department of Physics, KTH Royal Institute of Technology, Stockholm SE-10691, Sweden}
\begin{abstract}
 We consider a new class of topological defects in chiral magnetic crystals such
as FeGe and MnSi. These are composite topological defects that arise when skyrmions in the magnetic order intersect with twin boundaries in the underlying crystalline lattice.
We show that the resulting  stable configurations 
are a new type of defect that can be viewed as half-hopfions.
\end{abstract}
\maketitle

{\it Introduction} Topological defects and solitons are ubiquitous in nature and are responsible for many properties of materials. 
In condensed matter systems, there are two especially large categories of topological objects. One consists of those that form in classical field descriptions of the matter: i.e. topological defects or solitons in order parameter spaces such as vortices in superconductors, and domain walls and skyrmions in magnets and ferroelectrics \cite{chaikin1995principles,manton2004topological,svistunov2015superfluid}. 
The second category consists of topological defects in crystalline structures, i.e. irregularities in periodic atomic arrangements such as dislocations \cite{chaikin1995principles}.
Often the order parameter description applies for coarse-grained fields at large length scales and is not  affected in a topological sense  by crystalline defects.
I.e. traditionally, topological defects or solitons are textures of fields, such as magnetization, which are characterized by homotopy and corresponding invariants: topological index or charge. Often homotopy groups are in one-to-one correspondence with cyclic groups and, accordingly, the invariant is an integer  regardless of the deformations of the field configuration.

In this paper, we discuss  {\it composite} topological defects that involve both an order parameter and a symmetry breaking of the crystal lattice in the form of a twin boundary.

We focus on  chiral magnetic crystals with twinned boundaries.  
Certain chiral crystals, with the point-symmetry group $T$ are known to host skyrmions whose chirality depends on the chirality of the crystal \cite{bogdanov1994thermodynamically,MnSi2,FeGe,yu2010real,park2014observation,tokura2020magnetic}. 
We predict novel topological magnetic states involving skyrmions penetrating a twin boundary between left- and right-handed crystal subsystems.

The most interesting property of these composite defects that 
we find is that for the magnetization field they exhibit a fractionalization of topological charge  (in the common notation): i.e. they can be viewed as half-hopfion (which would be an unstable configuration in an ordinary magnet).
This generalizes the concept of fractionalization  of topological defects that currently attracts  significant interest in superconductors \cite{iguchi2023superconducting}, conventional skyrmions in ferromagnetic systems \cite{gao2020fractional,nagase2021observation},   boundaries of a Heusler materials \cite{jena2022observation}. 
and mathematical physics
\cite{jaykka2012broken,samoilenka2017fractional}. Lattices of fractionalized nuclear skyrmions were also recently proposed in a multicomponent superfluid \cite{samoilenka2020synthetic}.

\begin{figure*}[tb]
    \centering
    \includegraphics[width=19cm]{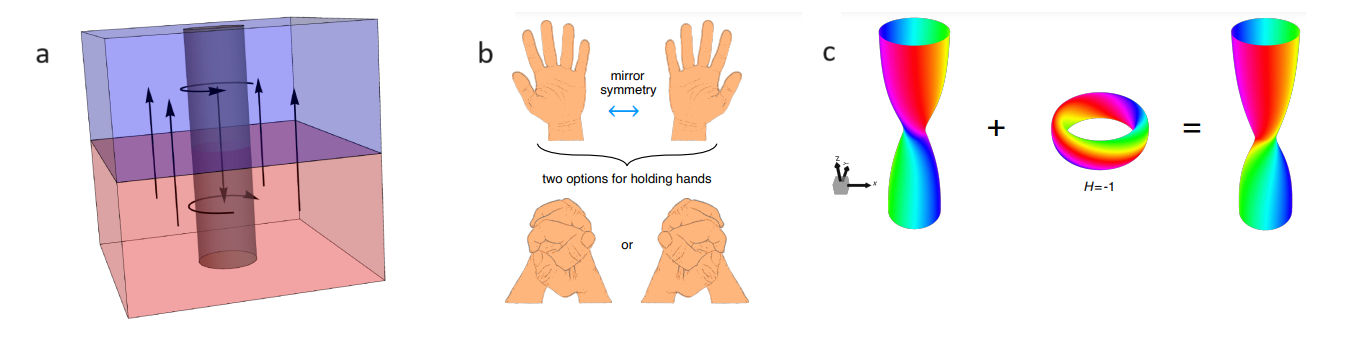}
    
    \caption{(a) A sketch of the composite model. Red and blue indicate the opposite lattice chiralities, while arrows indicate magnetization. (b) An example showing how two chiral objects, which are mirror images of each other, can join in two different ways to form chiral composite objects. (c) A magneto-crystalline defect  of one chirality can be combined with a hopfion to give a composite skyrmion of the opposite chirality. Each such defect can be viewed as half-hopfion.}
    \label{fig:hands}
\end{figure*}

{\it The model} Typical chiral crystals supporting skyrmions include B20 materials such as FeGe\cite{FeGe} and  MnSi\cite{MnSi2}. These are binary compounds with inherently chiral crystal structures. The B20 structure has a cubic lattice with four atoms in each unit cell; it has point group $T$ and space group $P2_1 3 (198)$. Because of their inherently chiral crystal structure, they can have a Dzyaloshinskii-Moriya interaction (DMI) \cite{dzyaloshinsky1958thermodynamic,moriya1960anisotropic} which is of the form $\propto \mathcal{D}\,\textbf{m}\cdot(\nabla\times\textbf{m})$, where $D$ is micromagnetic constant of DMI and $\textbf{m}\equiv\textbf{m}(\textbf{r})$ is the unit field representing the magnetization order parameter. 
It manifests as a tendency of the magnetization to twist clockwise or counterclockwise, depending on the sign of the constant $\mathcal{D}$. 

We refer to a crystal with positive $\mathcal{D}$ as right handed and one with negative $\mathcal{D}$ as left handed.

If there is a sufficiently strong applied field, the Zeeman interaction dominates over the DMI, and the solution is simply a uniform magnetization in the direction of the applied field. However, there can be tubes in which the magnetization is opposite to the field -- skyrmion strings. At the boundaries of these tubes, the magnetization changes from antiparallel to parallel. It does not change abruptly, but rotates, according to the sign of the DMI constant. 
The cross-section of such a tube is a symmetrical Bloch-type skyrmion. 
In lower fields, including down to zero, the stability of the skyrmion tubes is preserved~\cite{PhysRevLett.115.117201}, while the magnetic ordering surrounding them has the type of a conical spiral due to the interplay between the exchange interaction and the DMI.

It has been observed that B20 materials can have merohedral twinning \cite{Nitish}, such that the two crystals are of opposite chirality, with atoms in the interface belonging to both crystals and having the same coordination number as the bulk atoms. 

Consider such a twinned crystal 

 hosting skyrmions. 
Let there be a left-handed skyrmion in one crystalline subsystem and a right-handed skyrmion in the other, so that they continuously transform into each other at the twin boundary. Such a configuration is sketched in Fig~\ref{fig:hands}a.

{\it Symmetries of the model} Consider a simplified model with a left-handed subsystem at $z<0$ and a right-handed subsystem at $z>0$. 

Each of the two crystalline subsystems locally has a cubic structure, i.e. $T$ or $O$.  Globally, these symmetry groups are no longer possible, but nevertheless, it is possible to identify maximal subgroups from orthogonal group $O(3)$ that retain some symmetries. Such symmetries include those for which the detailed arrangement of atoms at the boundary is not taken into account, and only the domains outside the small boundary region $|z|>\delta$ are significant. In this way we can identify a point symmetry group for the system, neglecting the specific structure at the twin boundary. 
For example, in the case when the twinning plane coincides with $(100)$-type planes of crystalline subsystems we may have $C_{2h}$ or $S_4$ point groups if each crystal has $T$, and $C_{4h}$ if each crystal has $O$. The principal plane for them is the plane of the twin boundary.

{\it Micromagnetic Hamiltonian} Our analysis and calculations are based on the conventional micromagnetic Hamiltonian for cubic chiral magnets:
\begin{align} 
H\!=\!\int\!\mathrm{d}\mathbf{r}\ \Big(
\mathcal{A}\sum\limits_{i=x,y,z}\!|\nabla m_i|^2 
+ s(z)\,\mathcal{D}\,\mathbf{m}\!\cdot(\nabla\!\times\!\mathbf{m})
- \nonumber\\
- M_\text{s}\,m_z\,B_\text{ext} \Big)
,
\label{Ham}
\end{align}
where $\mathbf{m}$ is the unit vector field, $\mathbf{m}(\mathbf{r})=\mathbf{M}(\mathbf{r})/M_\text{s}$, 
$\mathcal{A}$ is the exchange stiffness, 
$\mathcal{D}$ is the DMI constant, 
$M_\mathrm{s}$ is the saturation magnetization, and 
$B_\text{ext}$ is the magnitude of external magnetic field pointed in the $z$ direction. 
The function $s(z)$ takes only two values: $s=-1$ for $z<0$ and $s=1$ otherwise, and aims to describe a system consisting of crystals of both chiralities joined in the $z=0$ plane. 
For convenience, we further use the dimensionless units as in Ref.~\cite{PhysRevB.87.094424} so that the distances are expressed in units of the cone/helix period $L_\text{D}=4\pi\mathcal{A}/\mathcal{D}$ 
and the magnitude of the applied field -- in units of the saturation field 
$B_\text{D}=\mathcal{D}^2/(2\mathcal{A}M_\text{s})$.  

The ground state of this Hamiltonian has a uniform magnetization in the direction of the external field, when that field is above a critical value. When the external field is subcritical, there is a modulation of the magnetization along the $z$ axis: it has a non-zero $z$ component, while the transverse component rotates about the $z$ axis. It takes the form $\textbf{m} = (m_{\perp}\cos(z/z_0),m_{\perp}\sin(z/z_0),m_{\parallel})$.

An important point is that the Hamiltonian~(\ref{Ham}) is invariant under inversion $\textbf{m}(\textbf{r})\to\textbf{m}(\textbf{-r})$ (note that the magnetization $\textbf{m}$ is a pseudovector and invariant under inversion; the same applies for $B_{ext}$). Therefore, the Hamiltonian is also invariant under a reflection about the $x-y$ plane (which also preserves $B_ext$):
\begin{align}
\left( 
\begin{array}{c}
m_x(x,y,z)\\ 
m_y(x,y,z)\\ 
m_z(x,y,z)   
\end{array}
\right)
\quad\leftrightarrow\quad
\left( 
\begin{array}{c}
-m_x(x,y,-z)\\ 
-m_y(x,y,-z)\\ 
m_z(x,y,-z)   
\end{array}
\right).
\label{HamSym}
\end{align}

{\it The solution} The left and right-handed skyrmions transform into each other via a Neel skyrmion; the magnetization in the interior and exterior remains unchanged, but in the intermediate layer, where the magnetization is orthogonal to the external field, it rotates by $\pi$. It can rotate in two ways and this determines the polarity of the intermediate Neel skyrmion. If we consider one of these solutions, it is not totally invariant under the transformation in (\ref{HamSym}), (only the asymptotics are invariant). Therefore, the existence of one solution guarantees the existence of another solution, related to it by (\ref{HamSym}), with the same energy. 
Thus there is a basis for a kind of dichotomy, which has an analog for simple chiral systems such as human hands, see Fig.~\ref{fig:hands}b. 
To find both types of solutions, we used the direct energy minimization method for~(\ref{Ham}) starting with different initial guesses.

For calculations we used ``Excalibur'' software~\cite{excalibur}. 
The found solutions are illustrated by tube-shaped surfaces corresponding to isosurfaces $m_z=0$ and presented on the left and right sides of Fig.~\ref{fig:hands}c. 

\begin{figure}
    \centering
     
    \adjincludegraphics[width=17cm,,Clip={.4\width} {.4\height} {0.4\width} {.4\height}]{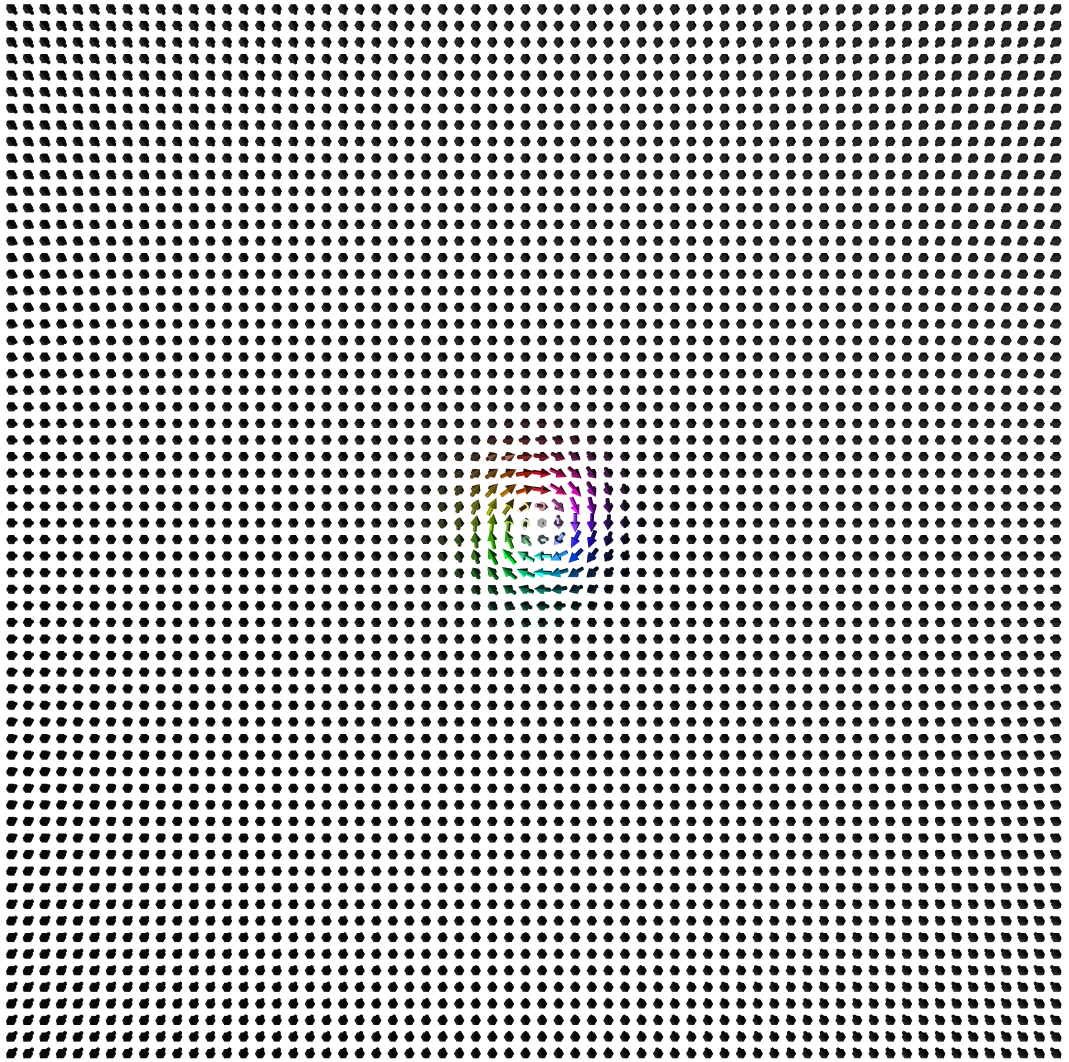}\hspace{0.01\textwidth} \adjincludegraphics[width=17cm,,Clip={.4\width} {.4\height} {0.4\width} {.4\height}]{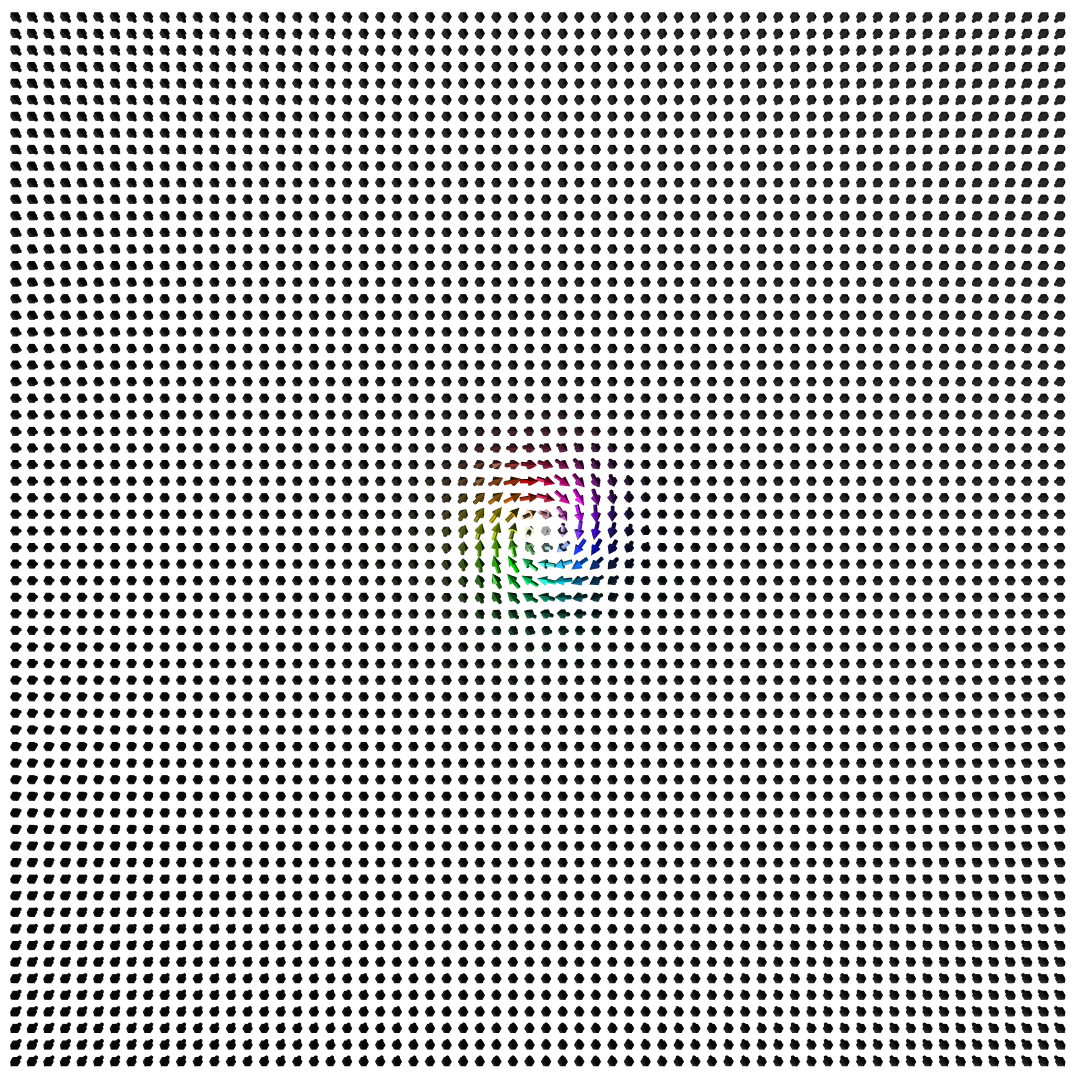}\vspace{0.01\textwidth}\\ \adjincludegraphics[width=17cm,,Clip={.4\width} {.4\height} {0.4\width} {.4\height}]{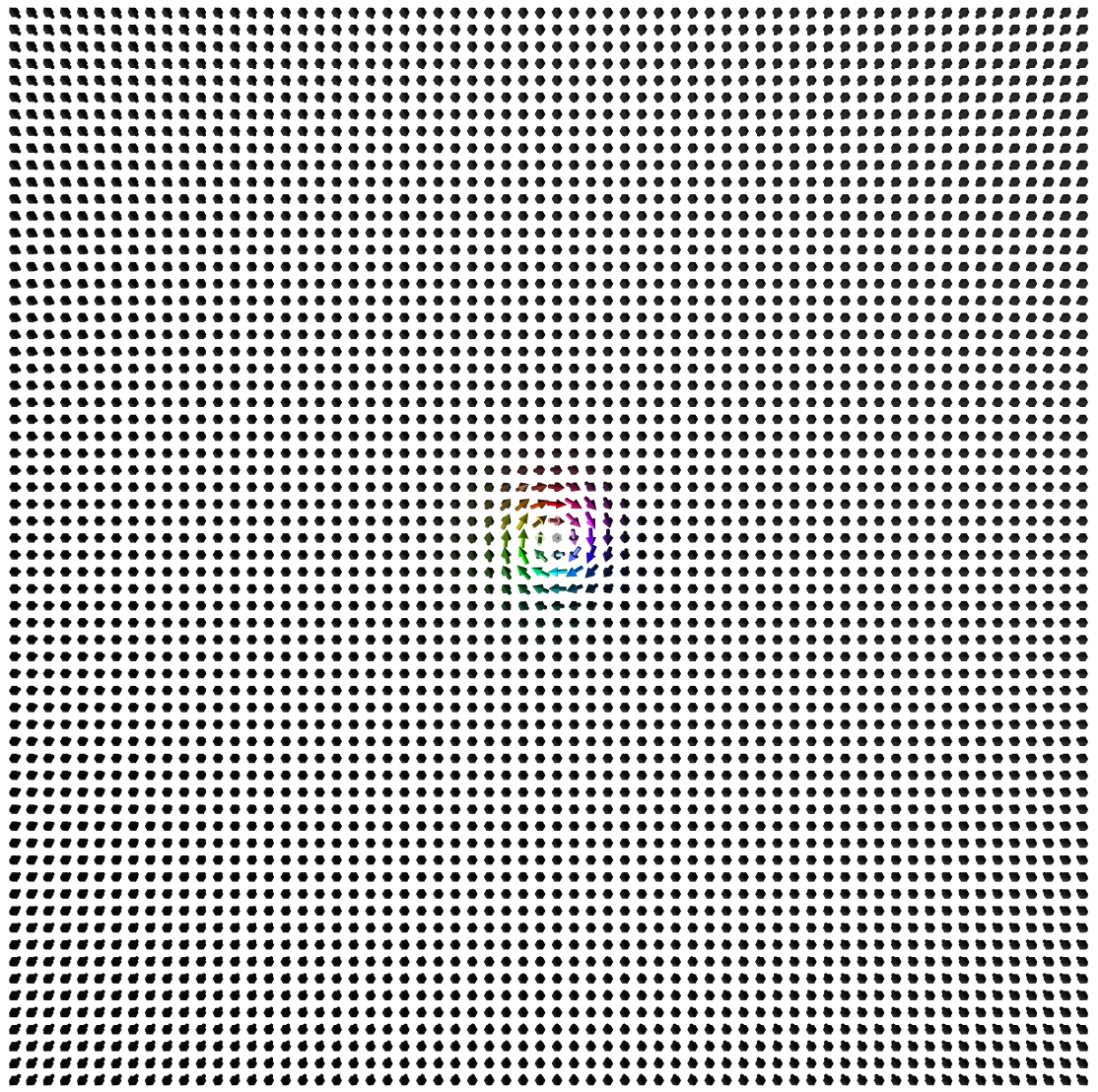}\hspace{0.01\textwidth} \adjincludegraphics[width=17cm,,Clip={.4\width} {.4\height} {0.4\width} {.4\height}]{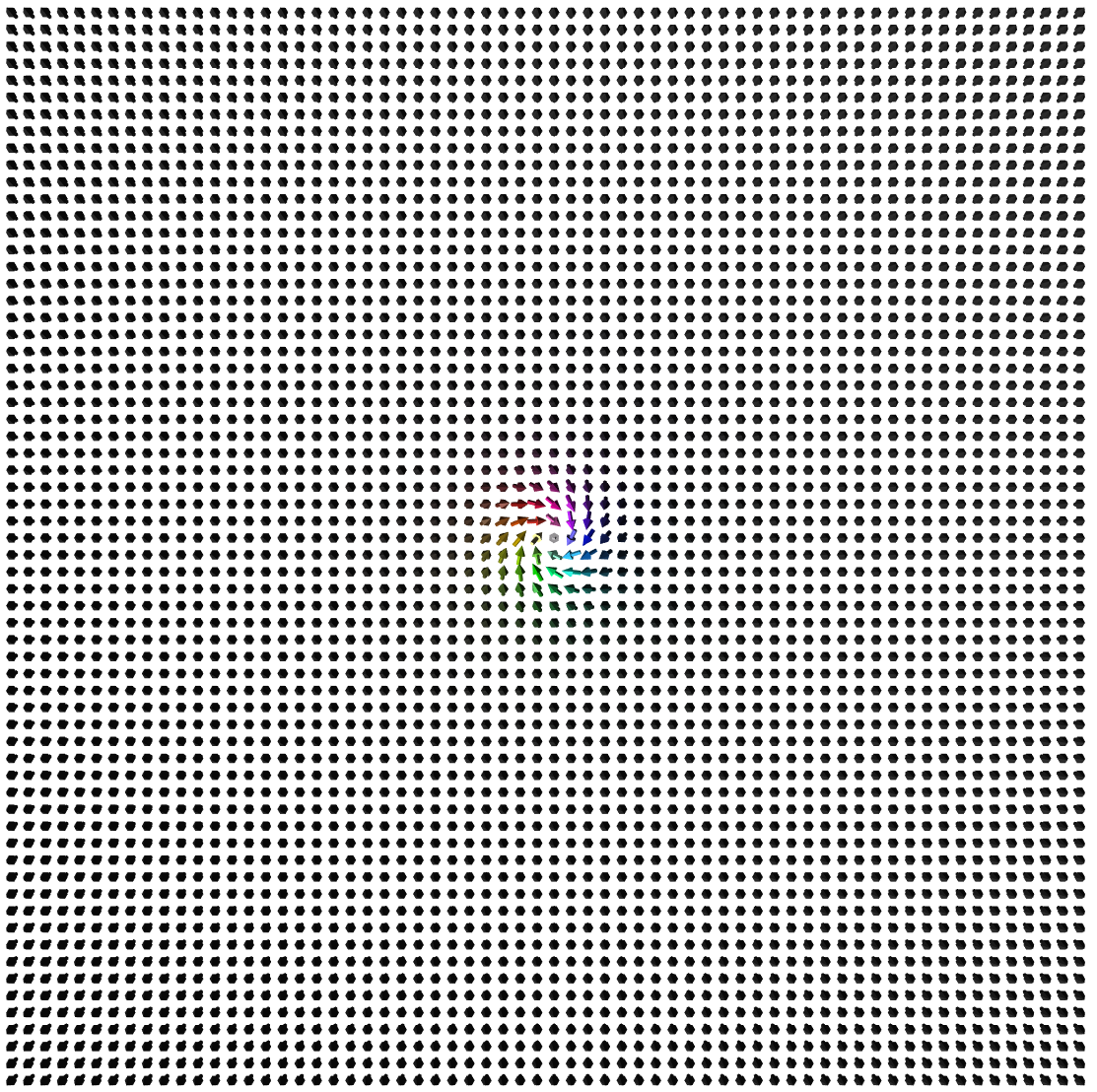}\vspace{0.01\textwidth}\\ \adjincludegraphics[width=17cm,,Clip={.4\width} {.4\height} {0.4\width} {.4\height}]{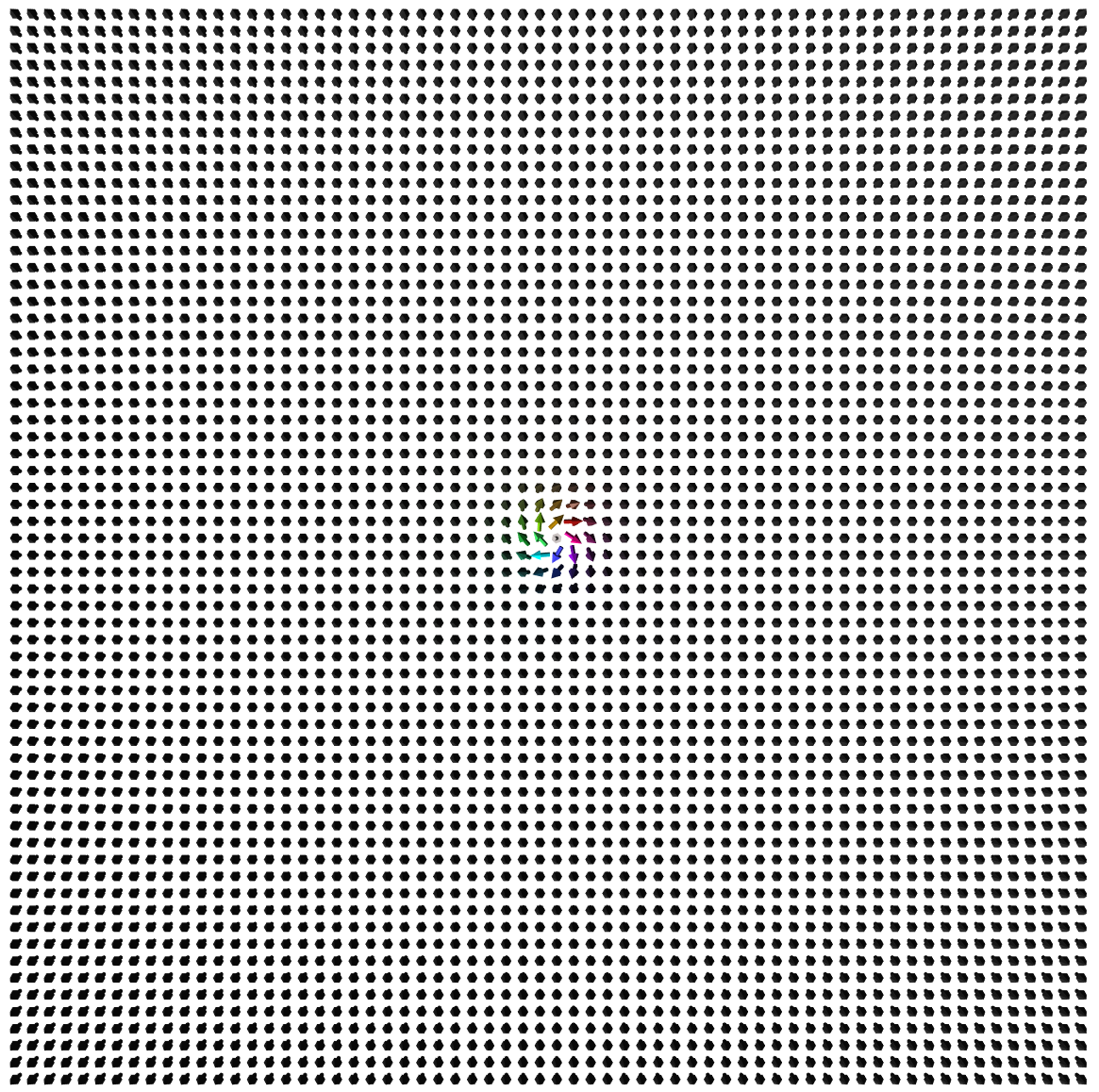}\hspace{0.01\textwidth} \adjincludegraphics[width=17cm,,Clip={.4\width} {.4\height} {0.4\width} {.4\height}]{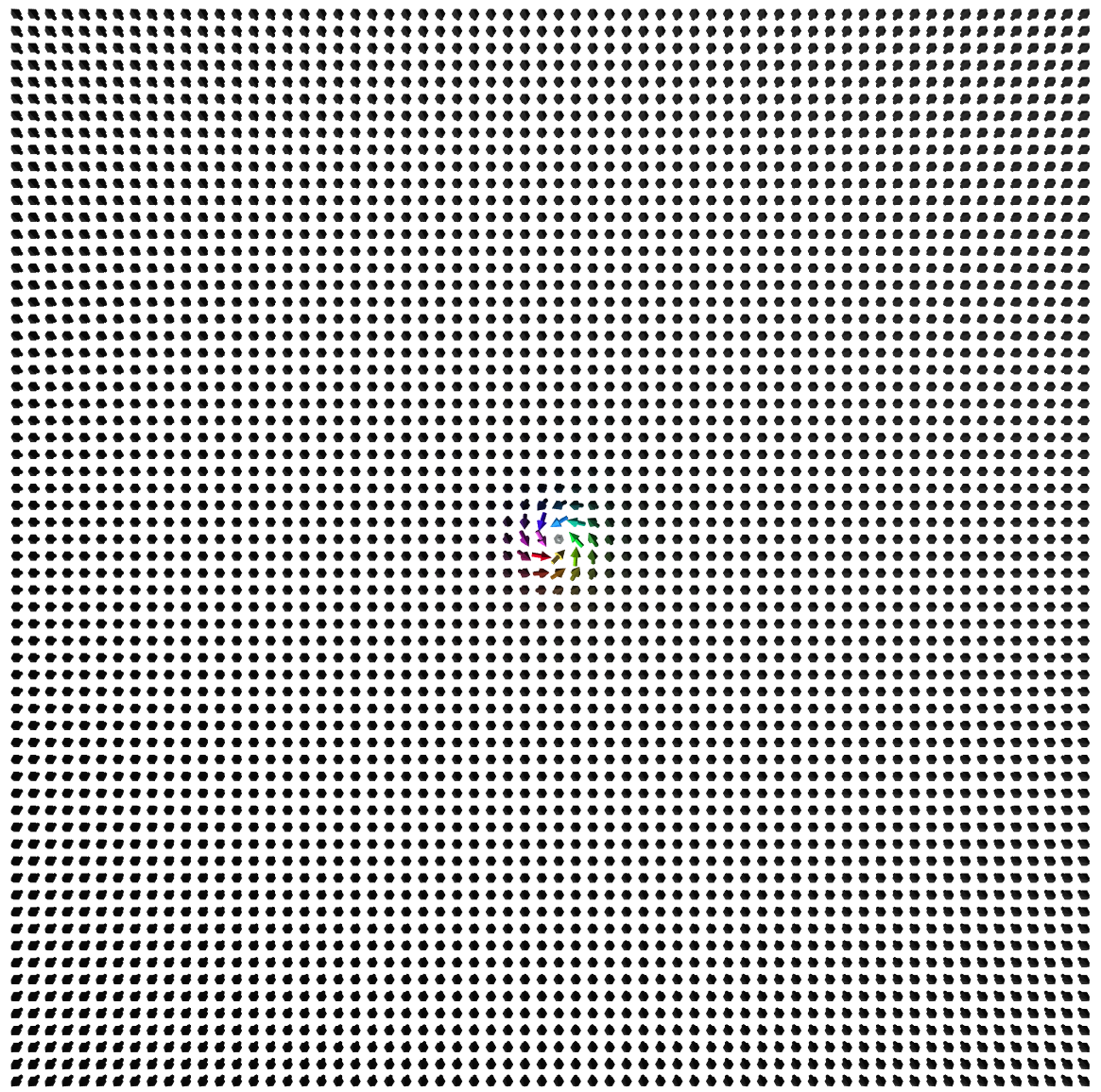}\vspace{0.01\textwidth}\\ \adjincludegraphics[width=17cm,,Clip={.4\width} {.4\height} {0.4\width} {.4\height}]{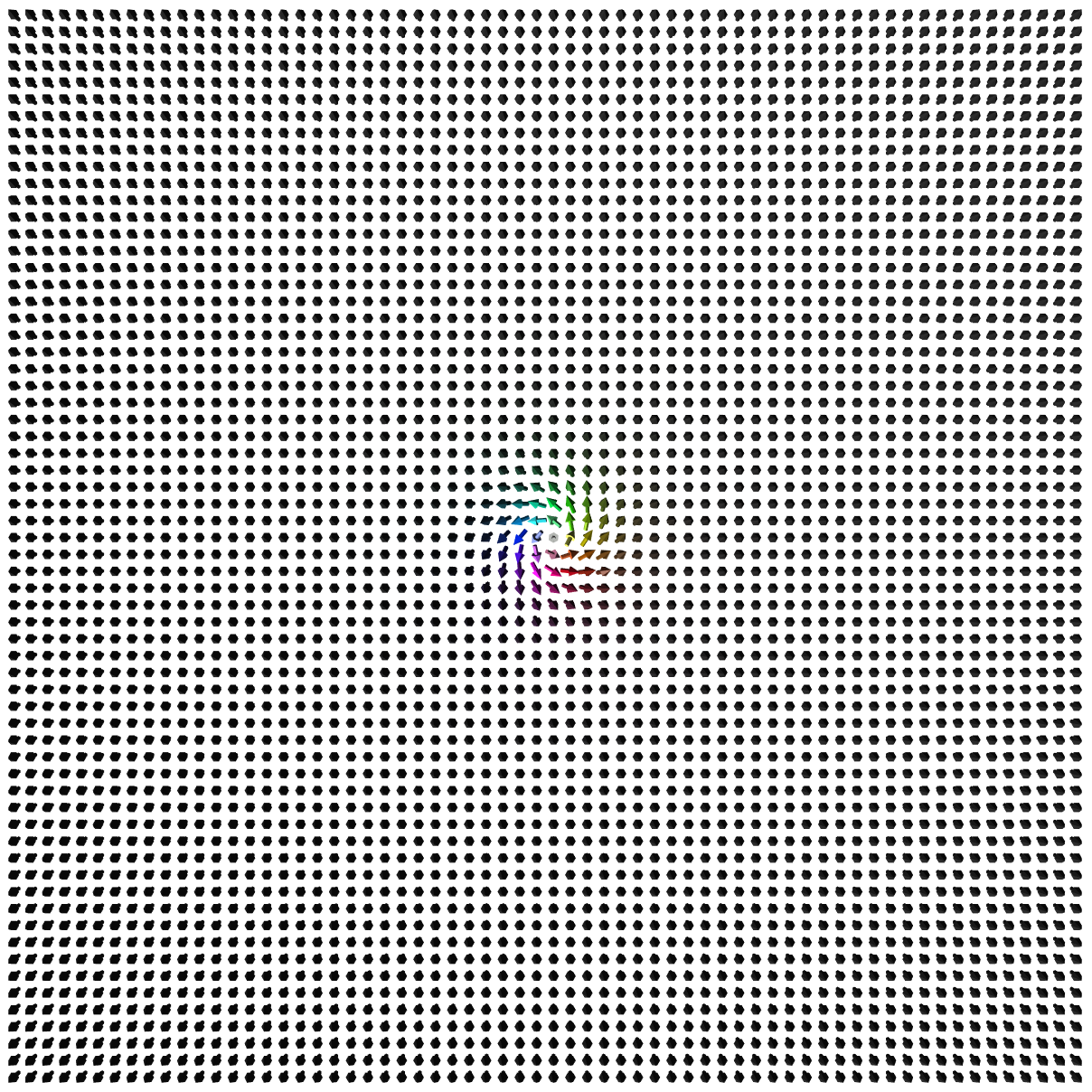}\hspace{0.01\textwidth} \adjincludegraphics[width=17cm,,Clip={.4\width} {.4\height} {0.4\width} {.4\height}]{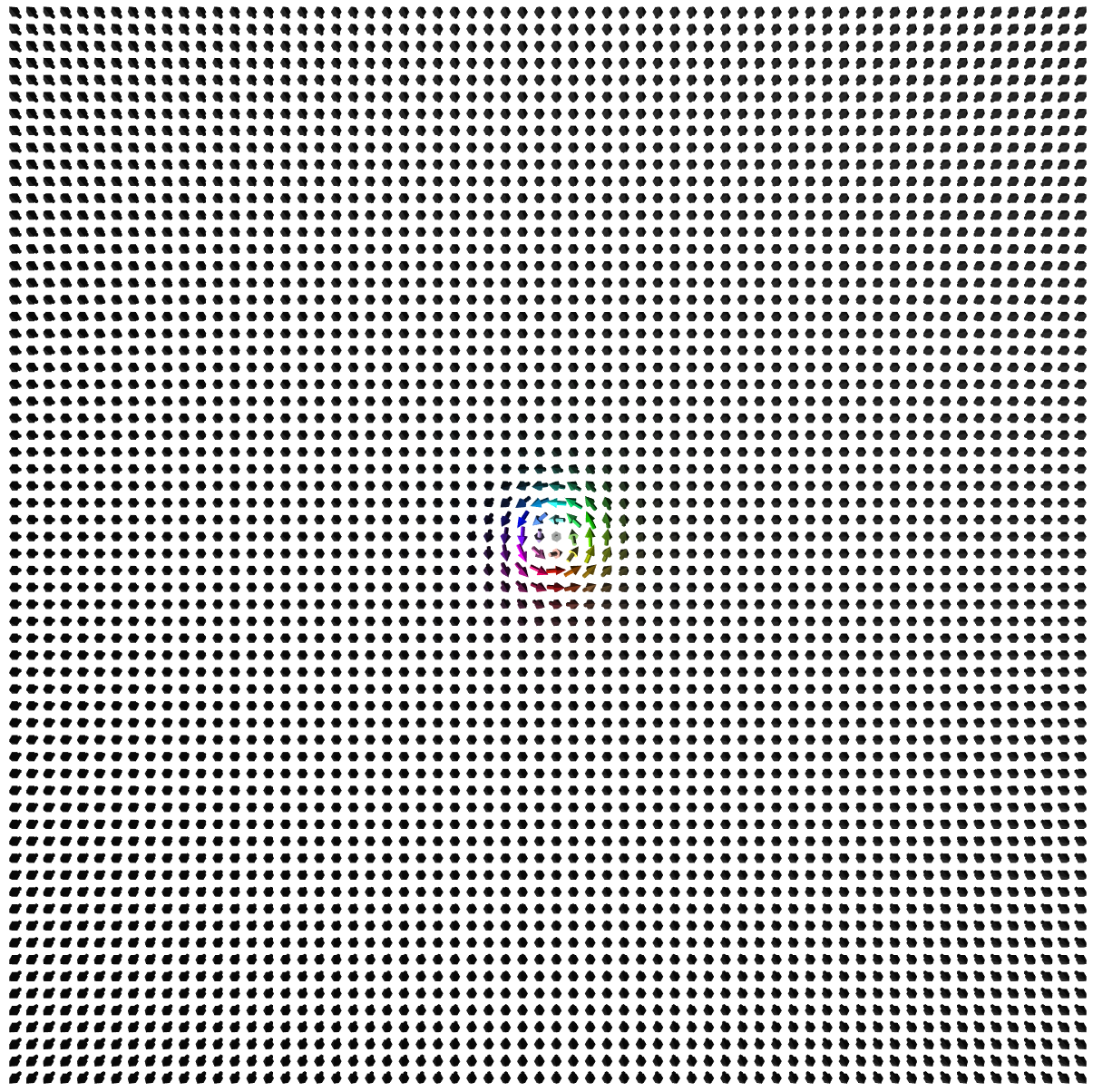}\vspace{0.01\textwidth}\\\adjincludegraphics[width=17cm,,Clip={.4\width} {.4\height} {0.4\width} {.4\height}]{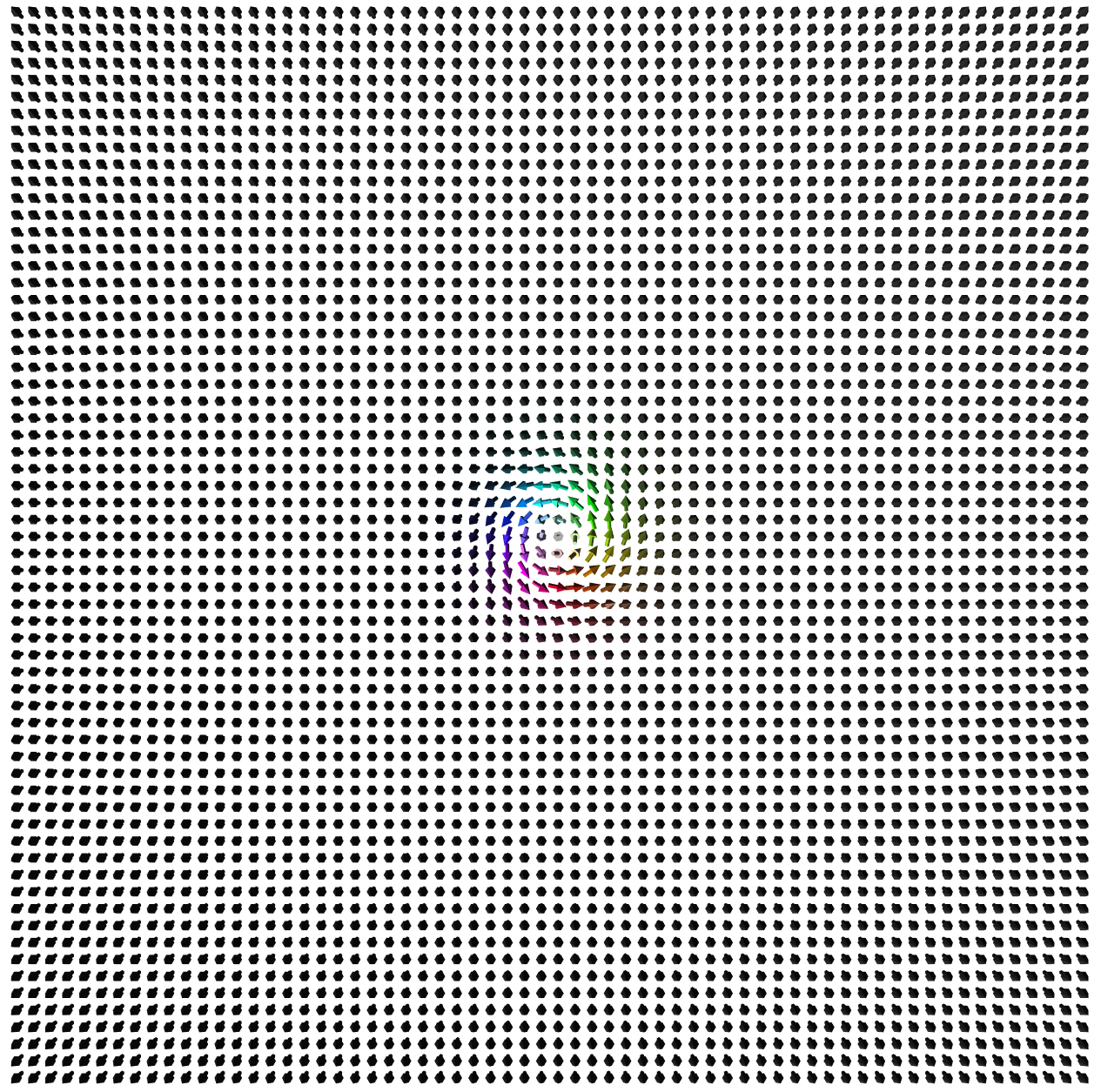}\hspace{0.01\textwidth} \adjincludegraphics[width=17cm,,Clip={.4\width} {.4\height} {0.4\width} {.4\height}]{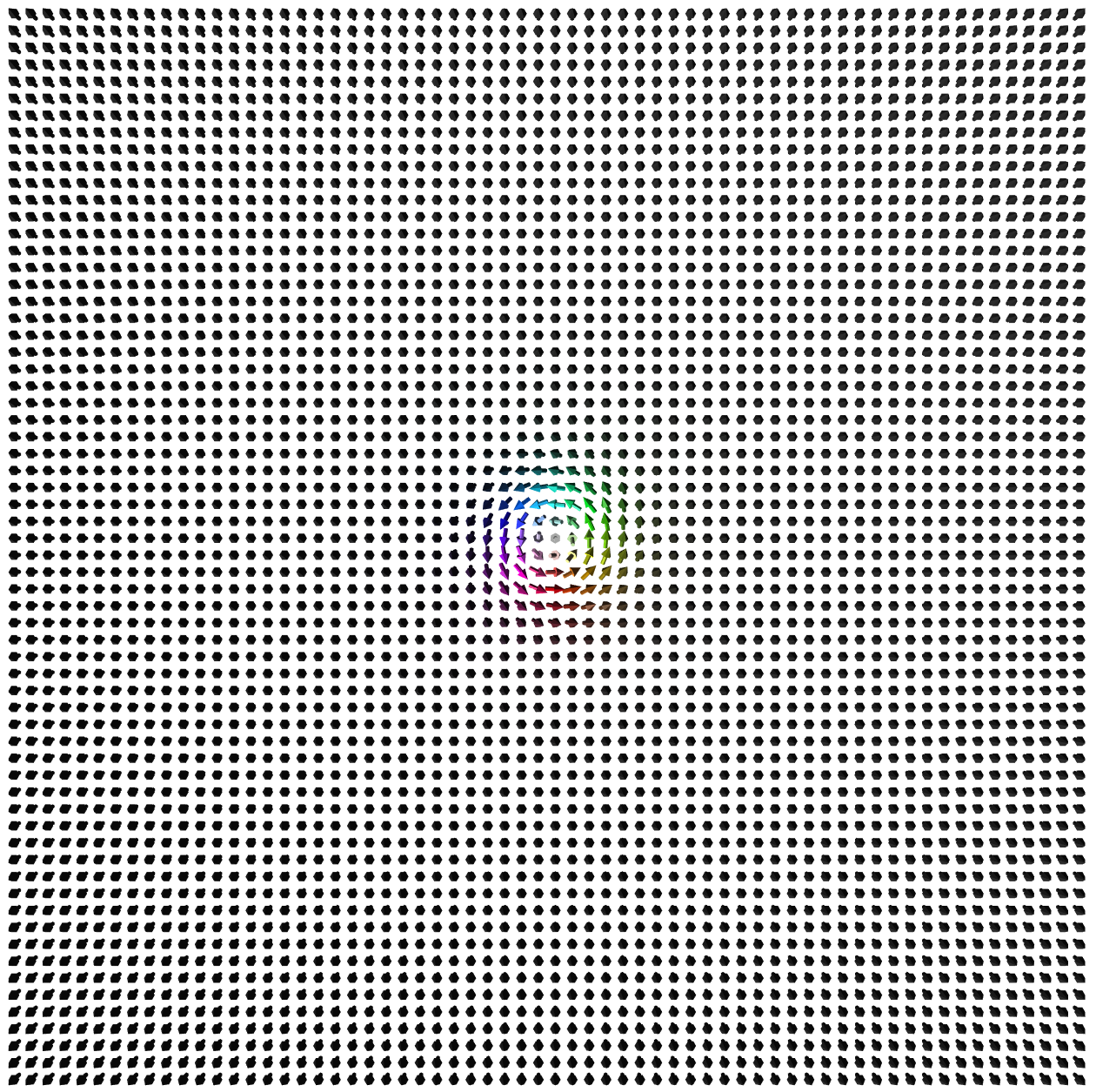}
    \caption{$x-y$ slices of the two solutions obtained in our numerical calculations. The arrows denote magnetic order parameter. The values of $z$ go from $-0.25,-0.125,0,0.125,0.25$ top to bottom. Black indicates positive $m_z$ and grey negative $m_z$.}
    \label{fig:horiz}
\end{figure}

\begin{figure}
    \centering
    \adjincludegraphics[width=16cm,,Clip={.38\width} {.38\height} {0.38\width} {.38\height}]{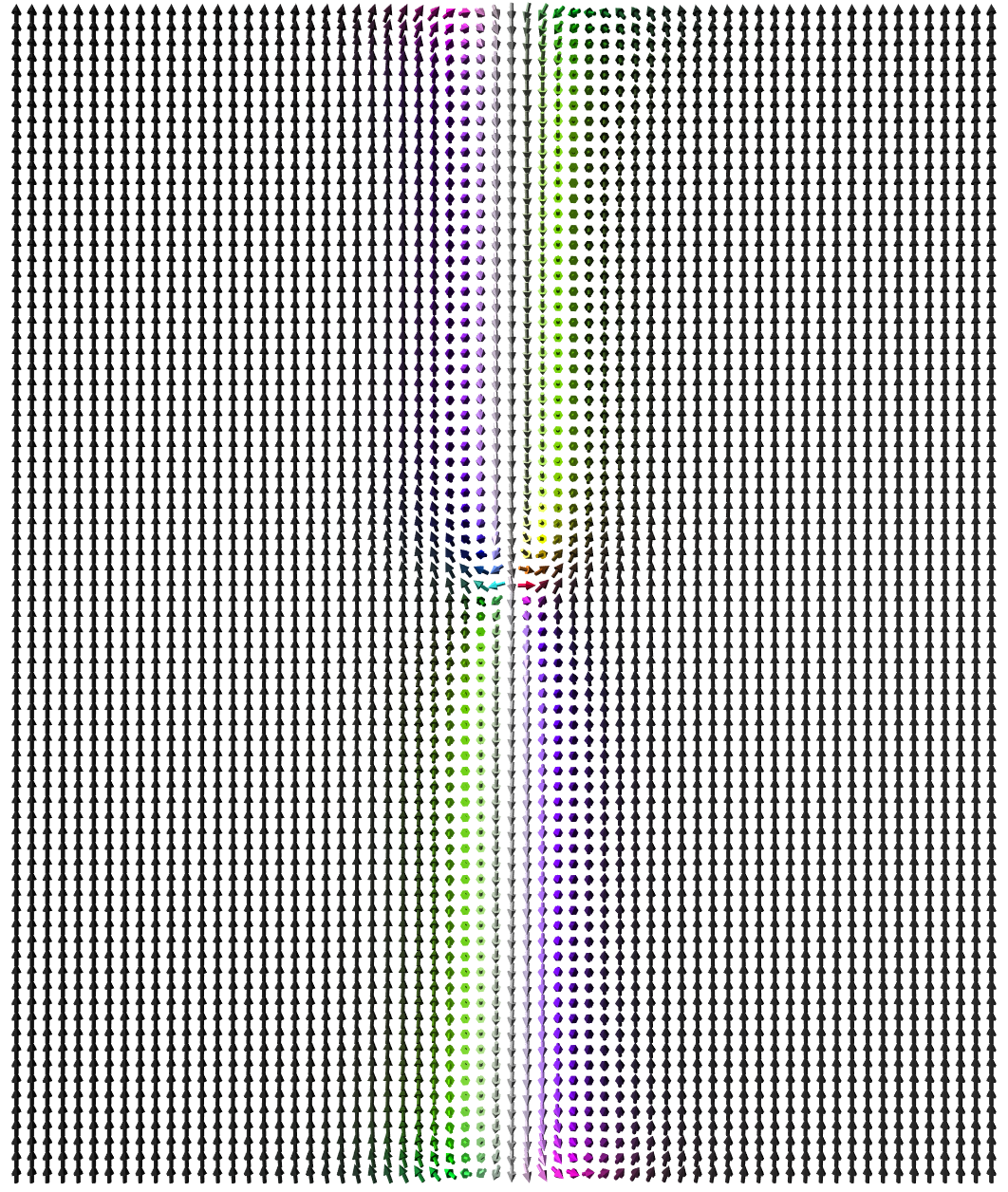}\hspace{0.01\textwidth} \adjincludegraphics[width=16cm,,Clip={.38\width} {.38\height} {0.38\width} {.38\height}]{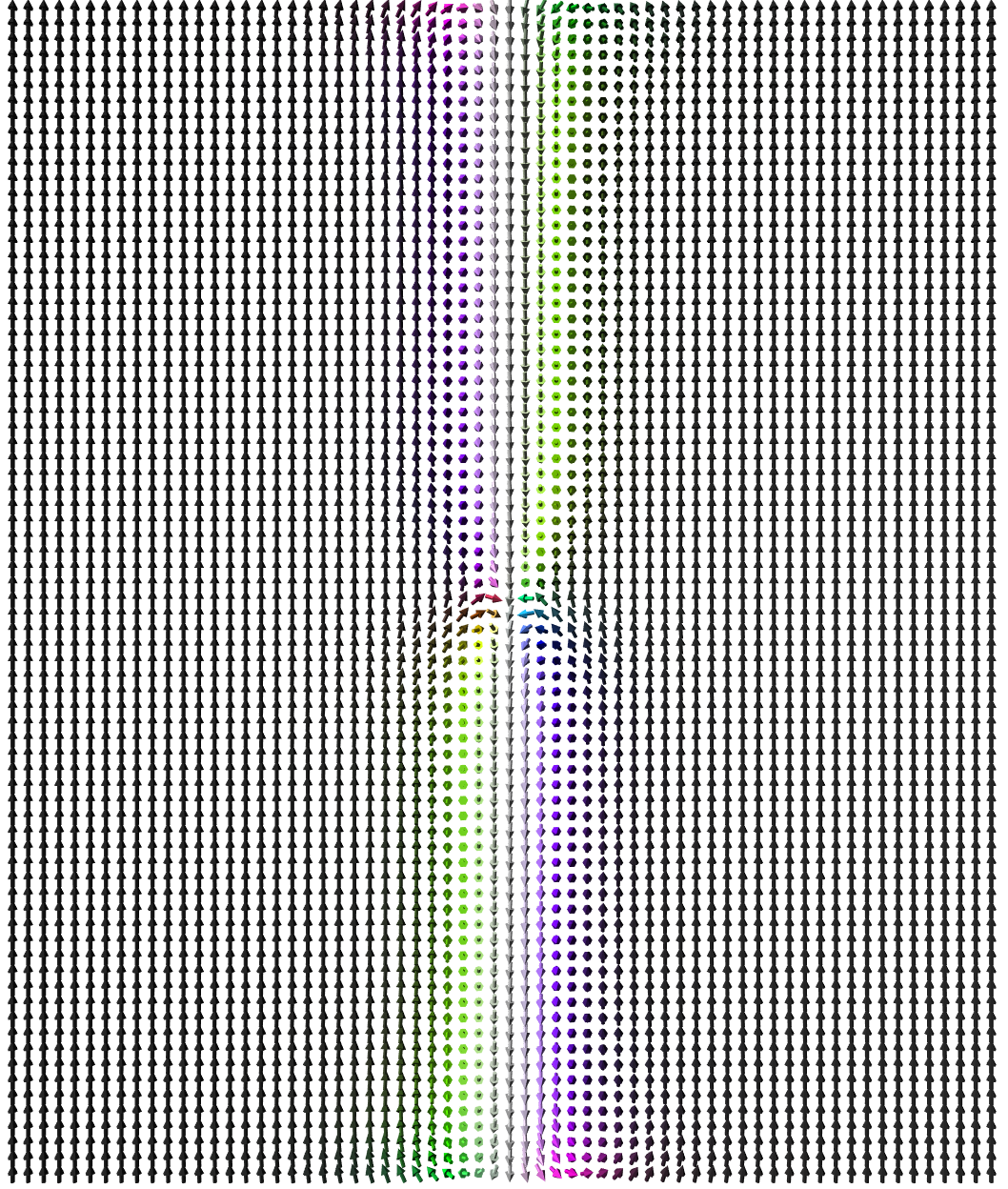}
    \caption{$y-z$ slices of the two solutions passing through the center. Note the transition from clockwise to counterclockwise does not happen exactly at $z=0$.}
    \label{fig:vert}
\end{figure}

We have shown $x-y$ slices of these solutions, from $z=-0.25$ to $z=0.25$ in steps of $0.125$ in Fig~\ref{fig:horiz} and $y-z$ slices in Fig~\ref{fig:vert}. Notice how the transition from a left handed to a right handed skyrmion does not take place exactly at the boundary. We consider a twin-boundary in the $x-y$ plane and magnetic field along $+z$. This means the magnetization is along $+z$ in the exterior and along $-z$ in the interior. Here, in the top (bottom) crystal, the DMI term 
 
is negative (positive), and as seen from above, the magnetization is oriented counterclockwise (clockwise). Now consider the transformation between these skyrmions of opposite chirality. In the solutions depicted in the left (right) halves of Figs~\ref{fig:horiz}~and~\ref{fig:vert}, as the $z$ coordinate increases, the magnetization rotates counterclockwise (clockwise), and the DMI term 
has a negative (positive) contribution. This is why the transition happens primarily in the top (bottom) crystal.

\begin{figure}
    \centering
    \adjincludegraphics[width=10cm,,Clip={.3\width} {.2\height} {0.3\width} {.2\height}]{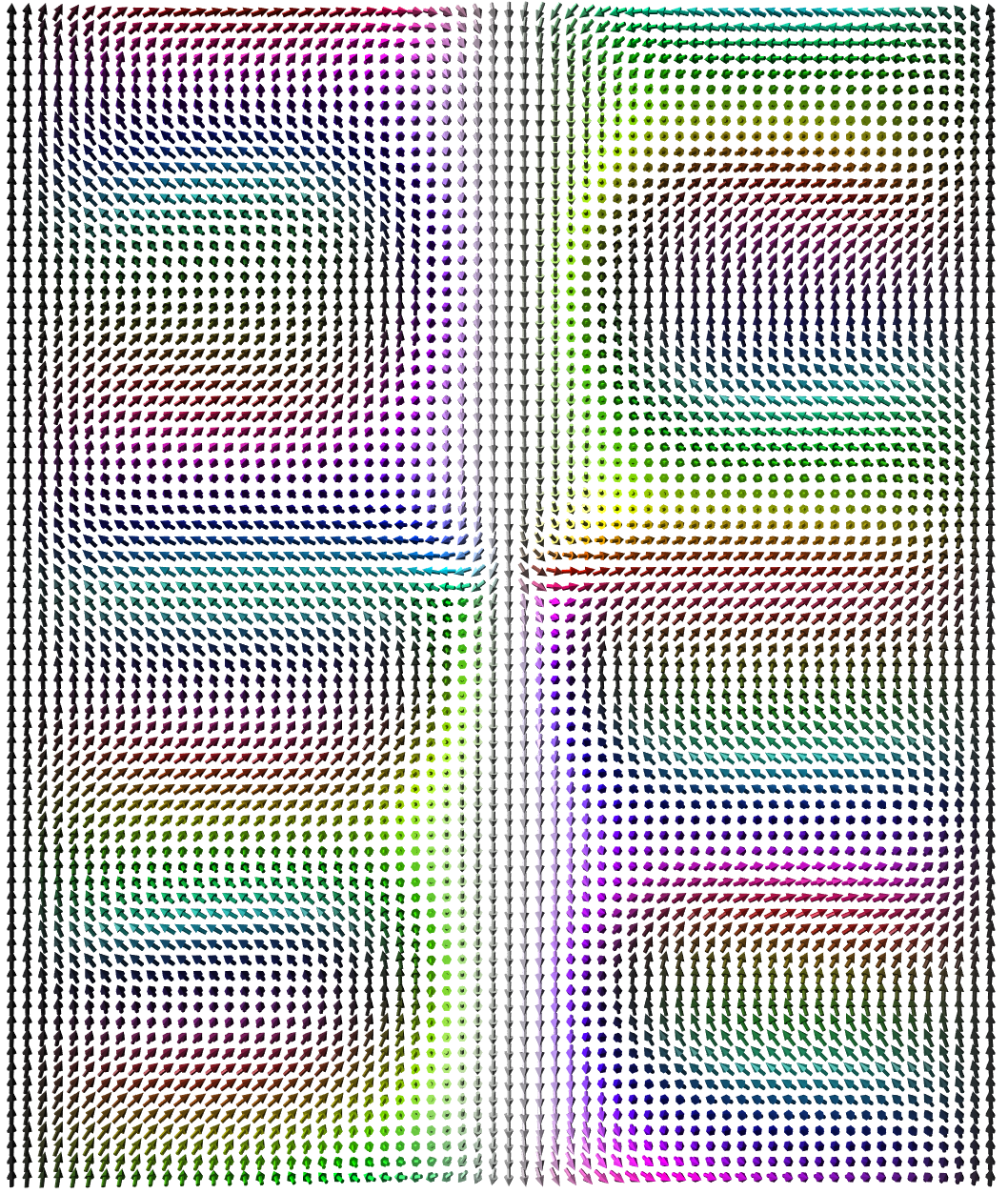}\hspace{0.01\textwidth} \adjincludegraphics[width=10cm,,Clip={.3\width} {.2\height} {0.3\width} {.2\height}]{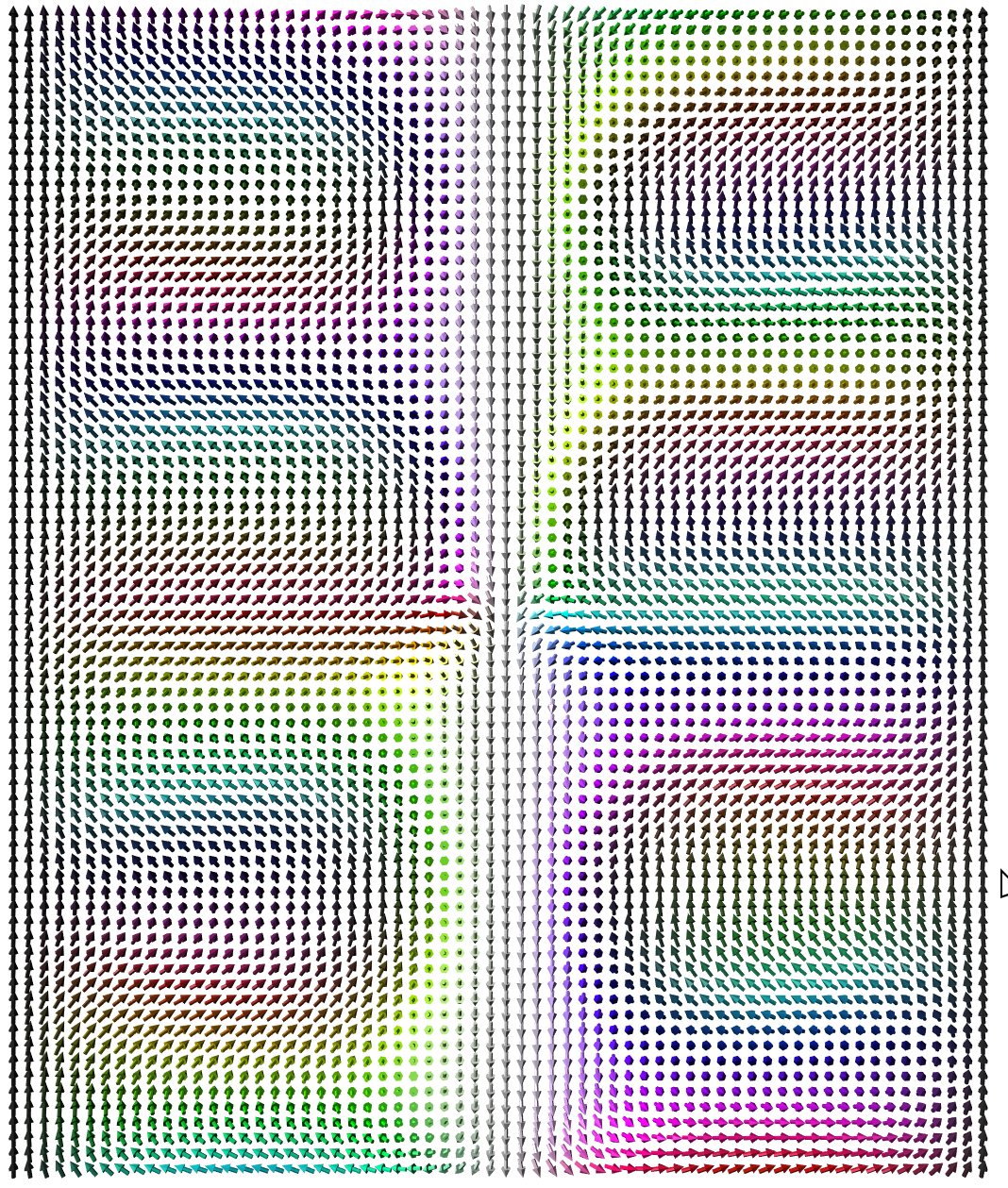}
    \caption{$y-z$ slices of the two solutions for a subcritical field, where background modulations are present.}
    \label{fig:helix}
\end{figure}

We have also plotted solutions for a less than critical external field in Fig~\ref{fig:helix} (i.e. when there is spontaneous modulation in magnetization). It can be seen that while the magnetization has a helical pattern outside the skyrmion, the skyrmion still remains, and the transition from clockwise to counterclockwise also persists.

 {\it Homotopy classification.} One obstacle to revealing the homotopy invariants of the found solutions is that the compactification approach in the considered system is not obvious. While in each plane $z=\text{const}$, the texture $\mathbf{m}(x,y)$ is localized on the homogeneous background and therefore is naturally compact, the field $\mathbf{m}$ nevertheless is different in the $x-y$ plane.
To get around this obstacle we use the auxiliary unit field $\mathbf{m}^\prime(\mathbf{r})$, which we define by the following continuous map of degree~2 from the spin sphere to itself: 
\begin{align}
\left( 
\begin{array}{c}
m_x\\ 
m_y\\ 
m_z   
\end{array}
\right)
\quad\rightarrow\quad
\left( 
\begin{array}{c}
m_x^2 - m_y^2\\ 
2 m_x m_y\\ 
m_z \sqrt{1 + m_x^2 + m_y^2}
\end{array}
\right) = \mathbf{m}^\prime.
\label{Smap}
\end{align}
Obviously, if the field $\mathbf{m}(\mathbf{r})$ is continuous, then $\mathbf{m}^\prime(\mathbf{r})$ is necessarily continuous. Therefore, the homotopy classes of field $\mathbf{m}^\prime$ are automatically inherited by the field $\mathbf{m}$. The advantage of $\mathbf{m}^\prime$ is that it is invariant under the transition~(\ref{HamSym}) and therefore $\mathbf{m}^\prime\big{|}_{z\rightarrow-\infty} = \mathbf{m}^\prime\big{|}_{z\rightarrow+\infty}$, meaning that $\mathbf{m}^\prime$ is presented in a compact domain. Using the method proposed in~\cite{Zheng_Kiselev_Rybakov_2023} we find that $\mathbf{m}^\prime$ is characterized by a pair of integer indices, $(Q^\prime,H^\prime)$ corresponding to the skyrmion and hopfion topological charges, respectively. 
By calculating the above indices using the method from~\cite{Zheng_Kiselev_Rybakov_2023} we find that $(Q^\prime,H^\prime)$ = $(-2, 2)$ and $(-2, -2)$ for the first and the second solutions, respectively. Thus, due to different $H^\prime$, these solutions belong to different homotopy classes. 
The corresponding topological-algebraic connection between our solutions for $\mathbf{m}$ field is illustrated in Fig.\ref{fig:hands}c, i.e. the homotopy transformation from one tube to another one requires the absorption of an isolated hopfion having its own charge $H=-1$. 
Conventionally, we can attribute to solutions for $\mathbf{m}(\mathbf{r})$   half-integer Hopf charges, i.e. $\frac{1}{2}$ and $-\frac{1}{2}$, respectively.

{\it Discussion} We have shown the possibility of a new kind of topological defect, that involves defects in
both crystalline and order-prameter fields.
The defect forms in a crystal with twinning plane.
Away from the plane the system forms
a chiral skyrmion  which intersects the twinning plane.
The plane induces a smooth deformation 
of this defect into a skirmion of different chiralty . This composite defect is in itself chiral, even though its constituents are mirror images of each other. A right handed composite defect cannot be smoothly transformed into a left handed one.
 A especially interesting aspect of these defects is that they are stable configurations that can be viewed as carrying a half-integer Hopf charge.

Since our characterization of the composite defect is topological, such objects are possible even in non-cubic chiral magnets, where the DMI term will incorporate anisotropic components.

 We discussed that these can be realized in magnetic B20 materials such as FeGe and MnSi.

\begin{acknowledgments}
    SK thanks  Nitish Mathur and.
    SK and EB thank Chris Halcrow for helpful discussions.
    Nordita is funded in part by NordForsk, the Nordic council of ministers. FNR acknowledges support from the Swedish Research Council.
EB was supported by the Swedish Research Council Grants 2016-06122 and 2022-04763, by Olle Engkvists Stiftelse    and by the Wallenberg Initiative Materials Science
for Sustainability (WISE) funded by the Knut and Alice Wallenberg
Foundation.

\end{acknowledgments}

\bibliography{bibliography}

\end{document}